# MatterTune: An Integrated, User-Friendly Platform for Fine-Tuning Atomistic Foundation Models to Accelerate Materials Simulation and Discovery

Lingyu Kong[†], Nima Shoghi[†], Guoxiang Hu[§], Pan Li[‡], Victor Fung[†]

[†]School of Computational Science and Engineering, Georgia Institute of Technology
[‡]School of Electrical and Computer Engineering, Georgia Institute of Technology
[§]School of Materials Science and Engineering, Georgia Institute of Technology
Corresponding author: `victorfung@gatech.edu`

### Abstract

Geometric machine learning models such as graph neural networks have achieved remarkable success in recent years in chemical and materials science research for applications such as high-throughput virtual screening and atomistic simulations. The success of these models can be attributed to their ability to effectively learn latent representations of atomic structures directly from the training data. Conversely, this also results in high data requirements for these models, hindering their application to problems which are data sparse which are common in this domain. To address this limitation, there is a growing development in the area of pre-trained machine learning models which have learned general, fundamental, geometric relationships in atomistic data, and which can then be fine-tuned to much smaller application-specific datasets. In particular, models which are pre-trained on diverse, large-scale atomistic datasets have shown impressive generalizability and flexibility to downstream applications, and are increasingly referred to as atomistic foundation models. To leverage the untapped potential of these foundation models, we introduce MatterTune, a modular and extensible framework that provides advanced fine-tuning capabilities and seamless integration of atomistic foundation models into downstream materials informatics and simulation workflows, thereby lowering the barriers to adoption and facilitating diverse applications in materials science. In its current state, MatterTune supports a number of state-of-the-art foundation models such as ORB, MatterSim, JMP, and EqformerV2, and hosts a wide range of features including a modular and flexible design, distributed and customizable fine-tuning, broad support for downstream informatics tasks, and more.

## 1  Introduction

Geometric machine learning models, such as graph neural networks (GNNs), have had a revolutionary impact on machine learning for the chemical and materials science domains. These models represent a paradigm shift away from the extensive and often time-consuming feature engineering required in traditional informatics approaches [1, 2, 3] and toward data-driven representation learning, thereby enabling them to be broadly applicable to a wide range of applications ranging from chemical property prediction and screening, molecular dynamics simulations, to the inverse design of new materials, and more. This has led to, by all accounts, an explosive growth in recent years of studies utilizing GNNs for these aforementioned tasks trained on existing materials datasets[4, 5, 6, 7, 8, 9].

Almost all GNN models of this class operate on the general principle of taking the atomic identity and structure of a molecule or crystal as inputs, and mapping this geometric information to their corresponding property labels. In the case of GNNs, this information is encoded in the node and edge attributes of a graph, which are then processed through message passing operations to yield latent atom-level and system-level embeddings or representations. From this embedding, the property labels can then be obtained via non-message passing layers, commonly referred to as a readout function or an output head. Starting from seminal examples such as SchNet[10] and CGCNN[11], subsequent models have incorporated increasingly sophisticated advancements including the incorporation of many-body interactions[12, 13, 14] equivariant features[15, 16, 17], and transformer-like architectures[18, 19], though nearly all still follow the same aforementioned general principles. Although these GNNs have become increasingly accurate and scalable with these improvements, they are inherently data driven and invariably function poorly for instances where training data is sparse. This limitation prevents their widespread application to the majority of materials science-related problems where data may range in the hundreds or even fewer samples.

A rapidly growing area of research towards greater data efficiency of GNNs is in the pre-training of GNNs. This approach generally involves first training these models on large upstream datasets (the "pre-training" stage) before continuing the training on the smaller downstream dataset(s) of interest (the "fine-tuning" stage). This process enables the models to learn robust, transferable representations without requiring the final property labels. Two general strategies exist for pre-training: supervised and unsupervised. In supervised pre-training, GNNs are initially trained on certain explicit property labels which are sufficiently generalizable to downstream needs. Properties such as energies, forces, and sometimes stresses from quantum mechanical calculations were found to be particularly effective for pre-training[20, 21, 22], among others[23, 24]. In unsupervised pre-training, unlabeled data are used instead, and the model is then trained on objectives such as a contrastive loss or denoising loss[25, 26, 27, 28]. While pre-training can be applied to datasets of any size and complexity, including artificial ones, there is a growing effort to pre-train GNNs on datasets which attempt to cover the full range of the chemical and materials space. Once pre-trained, these models should, in theory, be generalizable to downstream datasets of arbitrary complexity and properties. These models we term as "atomistic foundation models (FMs)." A growing numbers of studies have shown atomistic FMs can improve accuracies of GNNs significantly over models trained from scratch (i.e. without pre-training), as well as reduce data requirements by an order of magnitude or more[20, 21, 22].

Here, it is important to note the parallel development of universal interatomic potentials (UIPs), which are models trained to be broadly applicable force fields for systems of arbitrary complexity on compositions across the periodic table[29, 12, 30, 31, 28, 21, 32]. Whereas UIPs are intended to be used out-of-the-box for one specific task (as force fields), pre-trained models require an additional fine-tuning step before they can be used, but are applicable to tasks beyond force fields. Nevertheless, the distinction between UIPs and pre-trained models can become blurred as in some cases, the training procedures and datasets for UIPs can be identical to those used in the creation of pre-trained atomistic models, namely when the pre-training objective is on energies and forces.

Consequently, one can note that while not all pre-trained models can serve as UIPs, in general most UIPs should serve as capable pre-trained models.

Table 1: **Overview of Some Recently Released Atomistic Foundation Models**

| Model | Release Year | Num. Params | Dataset Size | Training Obj. |
| --- | --- | --- | --- | --- |
| MACE-MP-0 | 2023 | 4.69M | 1.58M | Energy, Forces, Stress |
| GNoME | 2023 | 16.2M | 16.2M | Energy, Forces |
| MACE-MPA-0 | 2024 | 9.06M | 12M | Energy, Forces, Stress |
| MatterSim-v1 | 2024 | 4.55M | 17M | Energy, Forces, Stress |
| ORB-v1 | 2024 | 25.2M | 32.1M | Denoising + Energy, Forces, Stress |
| JMP-S | 2024 | 30M | 120M | Energy, Forces |
| JMP-L | 2024 | 235M | 120M | Energy, Forces |
| eqV2-S | 2024 | 31.2M | 1.58M | Energy, Forces, Stress |
| eqV2-M | 2024 | 86.6M | 102M | Energy, Forces, Stress |
| DPA3-v1-MPtrj | 2025 | 3.37M | 1.58M | Energy, Forces |
| DPA3-v1-OpenLAM | 2025 | 8.18M | 143M | Energy, Forces |

Despite the demonstrated potential of atomistic FMs, general adoption by the broader scientific community is currently lacking, in large part due to the limitations of the available software infrastructure for its usage. While there is, to date, ample infrastructure for UIPs, this does not extend to any tasks beyond being used as force fields, such as materials property prediction. There is also limited standardization across different UIPs and atomistic FMs, resulting in a different package being needed for each different model, hampering benchmarking and workflow development. Finally, there is limited support for the customizability of the fine-tuning procedure, which is often hard-coded as a black-box method. As such, these existing packages do not currently fulfill the role of servicing atomistic FMs for general-purpose usage.

To address these limitations, we developed a modular, integrated, and user-friendly framework, called MatterTune, for fine-tuning atomistic foundation models to be applied to a broad range of materials science applications. The development of MatterTune follows several general design principles:

1. Highly generalizable and flexible abstractions that enable systematic extension while enforcing the necessary standardization.

2. Modular framework decoupling models, data, algorithms, and applications, enabling a high degree of adaptability and customizability for different materials informatics tasks.

3. Intuitive and user-friendly interfaces that simplify model fine-tuning and their application to downstream tasks.

So far, MatterTune has integrated several open-source atomistic FMs including JMP [20], ORB [28], EquiformerV2 [18], and MatterSim-V1 [21]. We fine-tuned these models using the MatterTune platform and evaluated them on representative materials informatics tasks, including molecular dynamics simulations, property screening, and materials discovery, demonstrating the performance and reliability of the MatterTune platform and its capabilities for data-efficient learning.

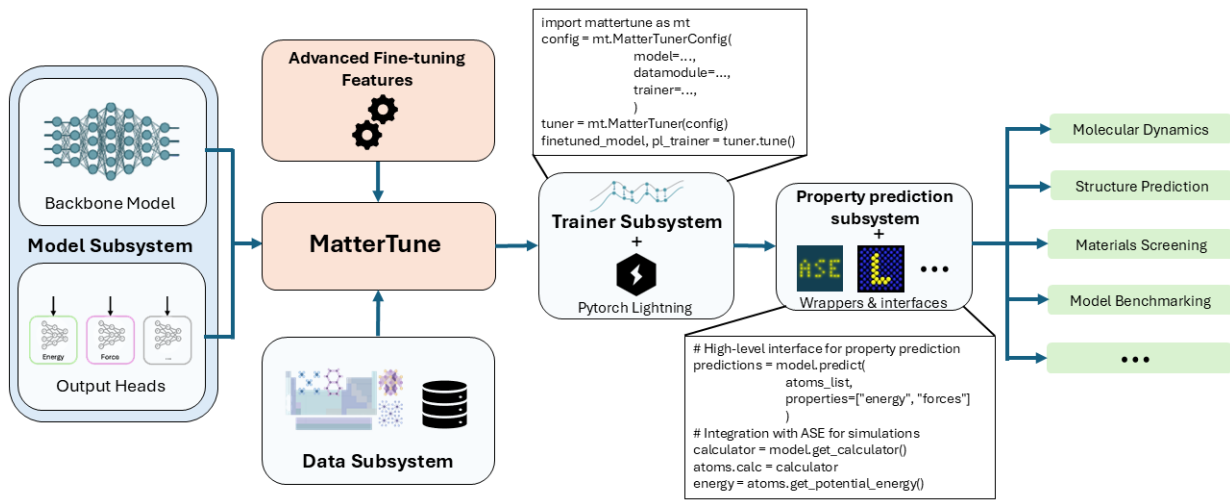

Figure 1: Overview of the MatterTune framework.

## 2 Methods

The MatterTune framework primarily consists of four components: a model subsystem, a data subsystem, a trainer subsystem, and an application subsystem, each covering a core component of a fine-tuning task. In addition, we have abstracted the key components that require standardization and extensibility, making it straightforward for MatterTune to quickly support new models and new data formats. The MatterTune package can be accessed at https://github.com/Fung-Lab/MatterTune.

### 2.1 Flexible and Generalizable Abstractions

In order to support various kinds of current and future atomistic FMs for the myriad of applicable materials systems and applications, some generalizable abstractions for atomistic FMs are needed. Several considerations need to be addressed here: First, atomistic FMs can employ diverse architectural paradigms, ranging from graph neural networks to transformers. Second, each atomistic FM has its own supported input format, internal representations, and computational requirements, all of which must be supported and standardized. Finally, the implementation should cover the breadth of possible materials informatics tasks and handle heterogeneous property types ranging from scalar quantities to vector fields. Considering these factors, MatterTune's architecture is centered around three key abstractions to enable systematic extension while enforcing the necessary standardization:

- **Data Abstraction**: The purpose of data abstraction is to provide unified support for as many input formats as possible for training and inference. We develop a minimal data abstraction that defines a dataset $\mathcal{D}$ as a mapping $f : \mathbb{N} \to \mathcal{S}$, where $\mathcal{S}$ represents the space of atomic structures in a standardized format. Given that different atomistic FMs require varying input formats, we choose `ase.Atoms` in ASE package [33] as the standardized format of $\mathcal{S}$. Individual atomistic FMs can then implement the necessary transformations from `ase.Atoms` to their respective input formats. Since the `ase.Atoms` format can store all structural and label information needed for training and prediction, this abstraction is broadly applicable.
- **Property Abstraction**: We introduce a property schema system that formally separates the

specification of physical properties from their model implementation, allowing users to focus solely on the types of properties they require from the model without concerning themselves with the details of how these properties are realized in FMs. This separation also enables the framework to handle both established properties like energy and forces as well as custom properties defined by users, and enforces type safety and physical constraints (e.g., energy conservation) in a property-specific manner.
- **Backbone Abstraction**: The purpose of backbone abstraction is to provide a set of unified functional interfaces for using different backbones, regardless of various FMs' completely different model architectures. For example, some key functions include the `model_forward` function, which handles forward propagation during prediction, and the `atoms_to_data` function, which converts input structures from the `ase.Atoms` format into the format required by the model. This abstraction ensures simplicity and consistency in model usage while enabling each model to retain its native internal representations and implementations.

## 2.2 A Modular and Standardized Framework Design

As illustrated in Figure 1, the modular framework enables MatterTune to decouple data, models, training algorithms, and downstream applications, allowing users to freely select and combine these components. This approach distinguishes MatterTune from other frameworks that rely solely on direct API calls. Following the aforementioned core abstractions, the framework is organized into several distinct subsystems:

- **The Data Subsystem** follows the aforementioned data abstraction and handles conversion between various materials science formats and a universal internal representation used by the MatterTune framework. Currently we have provided built-in support for common formats like XYZ, JSON, and ASE databases, which can be readily expanded to include additional formats as needed.
- **The Model Subsystem** is designed around the backbone and property abstractions, allowing users to simply specify the type of atomistic FM and the desired properties to predict in order to declare and construct a model. All implementation details—such as loading checkpoints, constructing output heads, handling input data, and performing forward passes, are automatically managed by MatterTune, respecting the original implementation of each atomistic FM. This approach enables users to leverage atomistic FMs without requiring in-depth knowledge of their underlying architecture and implementation.
- **The Trainer Subsystem** handles the general training, validation, and checkpointing of foundation models. A key design choice is made to integrate the training subsystem with PyTorch Lightning [34], a widely used and feature-rich training platform. This enables a range of critical capabilities while maintaining a clean separation of concerns between the model implementation and the training process. The integration of Lightning's abstractions allows MatterTune to maintain a modular and extensible architecture while still providing a simple, high-level interface for end users. Currently, MatterTune provides support for various optimizers and learning rate schedulers on the Lightning platform. It also includes implementations of data preprocessing statistics, exponential moving average, and other fine-tuning techniques, allowing users to select them freely. In addition, Lightning's callback features allow for ample flexibility for implementing more advanced fine-tuning strategies.
- **The Property Prediction Subsystem** provides the means for users to access the trained FMs

in an easy-to-use and intuitive manner to enable to quick integration of downstream materials tasks. This is accomplished by providing implementations of flexible wrapper classes for both general and targeted use-cases without having to deal with model architecture complexities or Lightning internals. As a starting point, we have implemented an `MatterTuneCalculator` which heritages the ASE [33] calculator interface, enabling direct use with established molecular dynamics and structure optimization algorithms available within the ASE package. For high-throughput material property prediction tasks, we have designed the `MatterTunePropertyPredictor` as a wrapper around atomistic FMs, enabling batch prediction in parallel. We are working on implementing interfaces for additional materials informatics simulation and computation software, such as LAMMPS.

## 2.3 A Unified Fine-tuning Technique Tool-kit

In the process of reproducing the fine-tuning experimental results for atomistic FMs, we discovered that reproducing their reported performance in the literature heavily depends on the specific fine-tuning approaches and techniques employed, and that different models require distinct fine-tuning settings. This observation motivated us to develop a unified fine-tuning toolkit on the MatterTune platform. Currently, our fine-tuning toolkit meets the requirement of ensuring that the integrated models can largely reproduce their expected fine-tuning performance as shown in Section 3. To this end, we have implemented the following:

- A variety of optimizers and learning rate schedulers. So far, MatterTune has supported the Adam [35], AdamW [36], and SGD [37] optimizers, as well as various learning rate schedulers including linear, step, exponential, cosine, and reduce-on-plateau. In addition to these, MatterTune also enables customization of optimizers and learning rate schedulers based on user needs. We support the application of different learning rates to different parts of the model, a technique that has been used and shown to be beneficial in the fine-tuning of models of the JMP series. Furthermore, we support combining multiple learning rate schedulers to achieve more sophisticated dynamic adjustments, such as cosine warm-up.
- Training generalization techniques such as Exponential Moving Average (EMA). Although the ablation studies in [20] suggest that EMA does not significantly improve fine-tuning performance on datasets such as MD17: Aspirin, MD22: Stachyose, QM9: $\Delta\epsilon$, MatBench: MP E Form, QMOF: Band Gap, and SPICE: Solvated Amino Acids, we believe these datasets are still not small enough in scale. In our experiments described in Section 3, where models are fine-tuned using only 30 data points, we observed that EMA actually helps improve both the stability and performance of fine-tuning.
- A comprehensive normalization system that handles both standard statistical normalization and physics-informed normalization schemes. Fine-tuning of foundation models (FMs) may involve multiple targets—for example, training a force field model typically involves three targets: energy, forces, and stress. Proper normalization helps balance the loss scales of these targets, ensuring that the training process converges more smoothly without being dominated by any single target. MatterTune currently supports not only standard normalization methods such as mean-std and root-mean-square, but also composition-based normalization using element-wise regression. Additionally, the normalization system is designed to be composable, allowing multiple normalization schemes to be applied in sequence.

# 3 Results

In the following experiments, we will demonstrate the performance of fine-tuned atomistic FMs on a variety of representative tasks and benchmarks using the MatterTune platform. These tasks include molecular dynamics simulations, materials property prediction, and materials discovery. The goal of these experiments is to showcase the correctness of MatterTune's implementation as well as its flexibility across diverse downstream applications.

We first note that MatterTune maintains strict adherence to the original implementations of each integrated atomistic FM. However, many models do not provide openly available details on fine-tuning parameters and techniques for specific tasks. Given that hyperparameter tuning is both complex and computationally expensive, we did not perform exhaustive hyperparameter optimization for the benchmarks shown below. As a result, we cannot guarantee that each atomistic FM achieves its best possible performance on these tasks. Nonetheless, for tasks with publicly accessible reference results, we have made a dedicated effort to reproduce them.

## 3.1 Property Prediction on MatBench and High-throughput Screening on Novel Out-of-distribution Materials

Atomistic FMs should be broadly applicable to various chemical and materials systems, and can be effectively fine-tuned as capable property predictors. This makes atomistic FMs highly promising for high-throughput property screening. In MatterTune, we provide support for constructing direct property prediction output heads for atomistic FMs, even if they are not present in their original implementations. To showcase this capability, we selected several foundation models for fine-tuning on multiple tasks from Matbench [38], a well-established materials informatics benchmark. The performance of JMP-S, ORB-V2, and Equiformer-31M-mp fine-tuned on Matbench is shown in Table 2. In our current tests, we perform fine-tuning on fold 0 of each dataset. In the table, we also list the fine-tuning performance on Matbench from the original JMP-S paper, as well as the best performance on the Matbench leaderboards. It should be noted that, since model fine-tuning can be a delicate process, variations in fine-tuning methods and hyperparameter configurations can lead to significant differences in the results. In our experiments, all models across all tasks were fine-tuned using the same configuration, so we cannot guarantee that the results reported on MatterTune represent the optimal performance of the models. Nonetheless, by comparing the fine-tuning results of JMP-S on MatterTune with those reported in the original paper, we found that we reproduced the reported accuracy in most tasks, with the only exception being formation energy, where our fine-tuning result was inferior to the original. Moreover, out of the three fine-tuned models JMP-S, ORB-V2, and Equiformer-31M-mp, the best model in each task significantly outperforms the current leading models trained from scratch on Matbench leaderboard.

Although Matbench provides a train-test split for evaluating fine-tuned models, they are drawn from the same original dataset distribution, which prevents them from accurately reflecting the models' performance on unseen new materials. To address this, we further performed high-throughput property predictions on approximately 381,000 new structures provided by the GNoME dataset [39] which are distinct from the original Materials Project dataset. For demonstrative purposes, we also screened out structures with band gaps between 1 eV and 3 eV and compared the classification performance with the ground truth. The results are shown in Table 3. The results indicate that the ORB-V2 model, which achieved the highest test accuracy on the band gap task in Matbench, also delivered the best performance in band gap property screening on the GNoME dataset.

Table 2: **Evaluation of Property Prediction Performance of Various Models on Matbench**

| Task(Units) | Best on Leaderboards (mean) | JMP-S-Baseline (fold0) | JMP-S (fold0) | ORB-V2 (fold0) | EqV2-31M-mp (fold0) |
|---|---|---|---|---|---|
| Dielectric (unitless) | 0.271 | 0.133 | 0.146 | 0.142 | **0.111** |
| JDFT2D (meV/atom) | 33.19 | 20.72 | **19.42** | 21.44 | 23.45 |
| Log GVRH ($\log_{10}$(GPA)) | 0.067 | 0.06 | 0.059 | **0.053** | 0.056 |
| Log KVRH ($\log_{10}$(GPA)) | 0.049 | 0.044 | **0.033** | 0.046 | 0.046 |
| MP E_form (meV/atom) | 0.017 | 13.6 | 25.2 | **9.4** | 24.5 |
| MP Gap (eV) | 0.156 | 0.119 | 0.119 | **0.093** | 0.098 |
| perovskites (eV/unitcell) | 0.027 | 0.029 | 0.029 | 0.033 | **0.026** |

Table 3: **High-throughput Screening on GNoME Band Gap Data**

| | JMP-S | ORB-V2 | Equiformer-31M-mp |
|---|---|---|---|
| MAE (eV) | 0.052 | **0.039** | 0.044 |
| Accuracy (%) | 98.16 | **98.80** | 98.53 |
| Recall (%) | 86.25 | **90.33** | 89.92 |
| F1 | 0.826 | **0.884** | 0.861 |

## 3.2 Few-shot Fine-tuning and Molecular Dynamics Simulations for Liquid Water

The original MatterSim paper presents a compelling demonstration of the capability of atomistic FMs to achieve reliable accuracy on specific tasks through fine-tuning with minimal data. In their experiments on a liquid water system, the authors compared three scenarios: zero-shot performance, fine-tuning using the full training set (900 samples), and fine-tuning with only a small subset (30 samples). The results showed that even just 30 samples for fine-tuning, the model achieved roughly the same level of accuracy as using 900 samples, and in both cases the models could accurately reproduce the radial distribution functions (RDFs) when compared with the experimental data.

To demonstrate the few-shot capability of the atomistic FMs in general, we followed the same experimental setup described in the original MatterSim paper. Out of the entire 1000 available ambient water data [40, 41, 42], we uniformly sampled 100 structures based on the energy distribution as a validation set and used the rest as the 900-sample dataset. We then randomly selected 30 structures from the 900-sample dataset and subsequently repeated them until a new dataset comprising 900 samples was obtained. We refer to this dataset as the 30-sample dataset. We fine-tuned various foundation models on both the 900-sample and the 30-sample dataset and evaluated models' mean absolute errors on the validation set. The results are shown in Table 4.

Table 4: **Fine-tuning Performance of Various Foundation Models on Ambient Water Dataset**

|  | MatterSim-V1-1M | | JMP-S | | ORB-V2 | | EqV2-31M-mp | |
| --- | --- | --- | --- | --- | --- | --- | --- | --- |
|  | 900 samples | 30 samples | 900 samples | 30 samples | 900 samples | 30 samples | 900 samples | 30 samples |
| $MAE_E$ (meV/atom) | 1.21 | **1.20** | 3.06 | 5.65 | 2.50 | 7.52 | 2.76 | 4.98 |
| $MAE_F$ (meV/Å) | 38.37 | 40.65 | **19.98** | 30.17 | 33.73 | 64.03 | 22.41 | 35.21 |

We further conducted 200-ps molecular dynamics (MD) simulations of a water structure with 192 atoms per unit cell at 298 K using the foundation models fine-tuned on the dataset of 30 samples shown above. The MD thermostat engine employs the NPT ensemble implemented in ASE (without external stress to keep the cell fixed). The results of the radial distribution function analysis are shown in 2. Interestingly, we observed that although all four models performed well in terms of MAE, as shown in Table 4, the MD simulation results varied significantly. Only MatterSim and EquiformerV2's results fit the experimental data well, whereas the RDF results for JMP-S and ORB-V2 exhibited clear problems. This observation echoes the statement in [43], which cautions that evaluating models solely based on force MAE can lead to misleading conclusions. One hypothesis is that JMP-S and ORB-V2 employ direct force prediction without ensuring energy conservation, while MatterSim's does maintain conservation, which might explain the differences in RDF outcomes. However, EquiformerV2 also uses direct force prediction yet still produces roughly correct RDF results.

## 3.3 Zero-shot Prediction and Structural Geometry Optimization

Novel material discovery and structure prediction are among the central challenges in the computational materials sciences. Matbench Discovery [46] provides a benchmark for evaluating models in accurately determining stable materials structures. We validated MatterTune's implementation of model loading and zero-shot prediction, as well as its correct support for geometry optimization, by reproducing the performance of several models on Matbench Discovery. We employed the ASE-implemented FIRE optimizer and the ExpCellFilter unit cell filter for all models, using 0.02 eV/Å and a maximum of 500 steps as the cut-off conditions for structural relaxation. The final results, along with a comparison to the leaderboard outcomes, are presented in Table 5. The results indicate that the reproduced outcomes are within an acceptable error range relative to the leaderboard results. The very minor discrepancies observed may stem from the choice of optimizer, the unit cell filter, and numerical precision, among other factors.

## 3.4 Representation Space Visualization

It is generally believed that one of the main reasons foundation models excel in downstream tasks is their ability to learn high-quality, general-purpose geometric representations during pre-training.

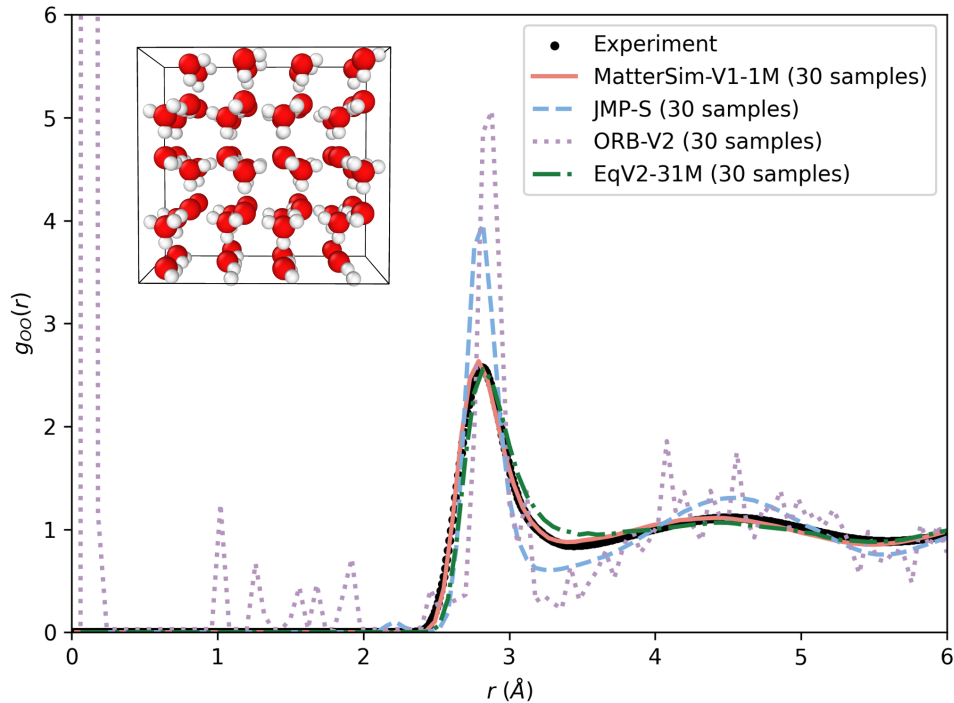

Figure 2: Oxygen-oxygen radial distribution functions of ambient water under 298K obtained from foundation model based MD simulations. Black dots represent experimental references from [44, 45]

Table 5: **Evaluation of Zero-Shot Performance of Various Models on Matbench-Discovery**

|  | EqV2 S DeNS Baseline | EqV2 S DeNS | MatterSimV1 5M Baseline | MatterSimV1 5M | ORB-V2 Baseline | ORB-V2 |
|---|---|---|---|---|---|---|
| F1 | 0.815 | 0.792 | 0.862 | 0.842 | 0.880 | 0.866 |
| DAF | 5.042 | 4.718 | 5.852 | 5.255 | 6.041 | 5.395 |
| Prec | 0.771 | 0.756 | 0.895 | 0.876 | 0.924 | 0.899 |
| Acc | 0.941 | 0.925 | 0.959 | 0.949 | 0.965 | 0.957 |
| MAE | 0.036 | 0.035 | 0.024 | 0.024 | 0.028 | 0.027 |
| R2 | 0.788 | 0.780 | 0.863 | 0.848 | 0.824 | 0.817 |

We leveraged MatterTune's internal feature extraction capabilities to export the node representations for four different models. To demonstrate this, we selected two datasets: MPTraj and MP_E_form, and visualize the representations generated by pre-trained foundation models and those fine-tuned for specific tasks. For the MatterSim, JMP, and ORB models, which intrinsically contain node feature vectors, we used these as the structural representations of atomic local environments. In Equiformer, where node features are divided into multiple irreducible representation channels, we extracted the invariant (i.e., $l = 0, m = 0$) features as a representative structural descriptor. Extracting the latent representations of atomistic FMs can provide a window into interpreting their performance, as well as be used for purposes such as active learning[47].

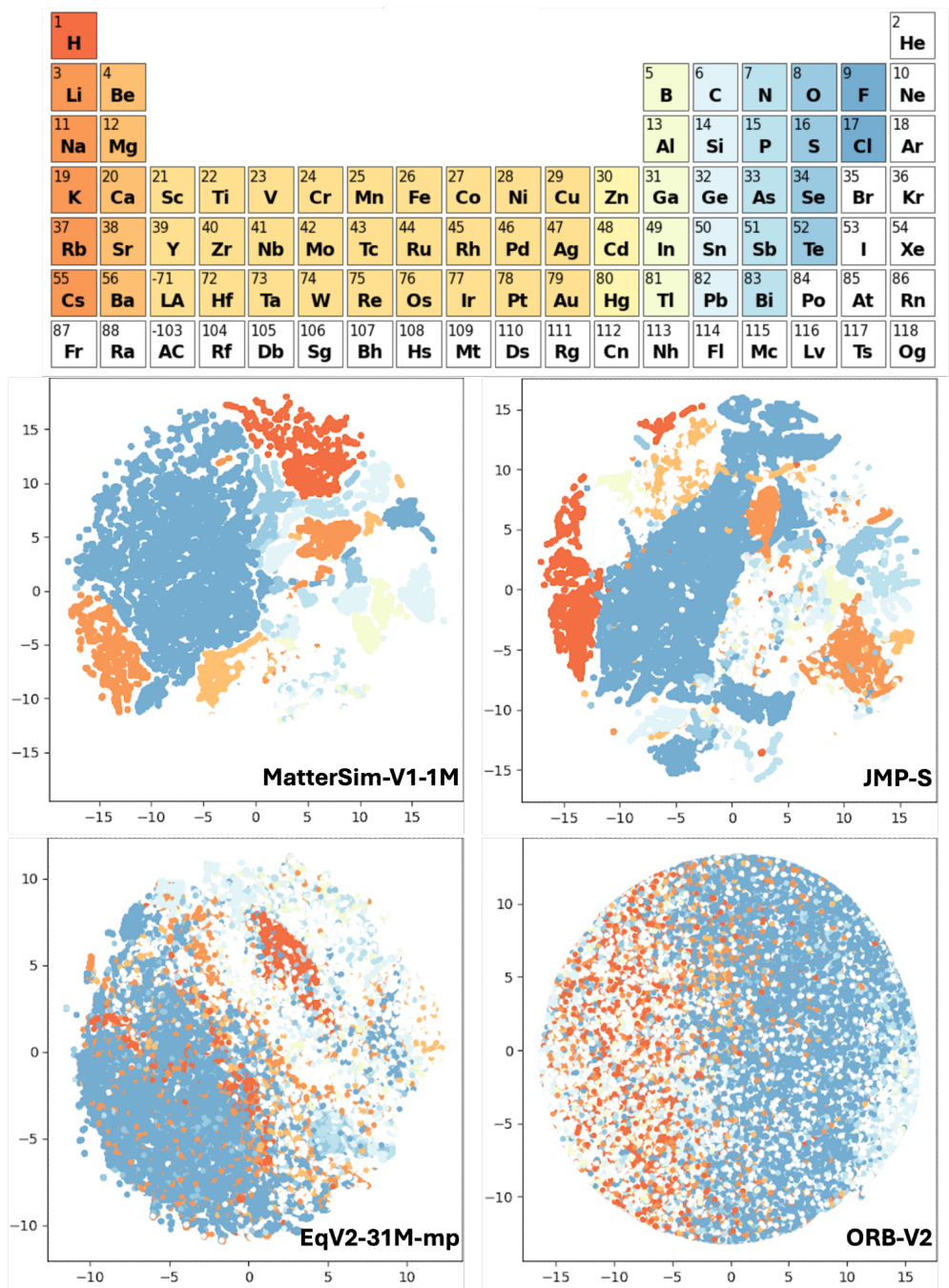

Figure 3: TSNE visualization of four pre-trained foundation models' node representations on a subset of MPTraj dataset containing 5000 structrues. Node representations are colored by element types, with similar colors assigned to elements in the same group.

In Figure 3, the representations of four different atomistic FMs are visualized using the t-SNE algorithm. The results show that the MatterSim and JMP models clearly capture the clustering of elements within the same group. This is to be expected from a chemical perspective, and suggests some level of transferable chemical knowledge is trained into these models. In contrast, the clustering for Equiformer and ORB models is less pronounced, especially for ORB. These results highlight the remarkable diversity in the internal representations of current atomistic FMs, which may arise due to differences in training objectives, training data, and model architectures.

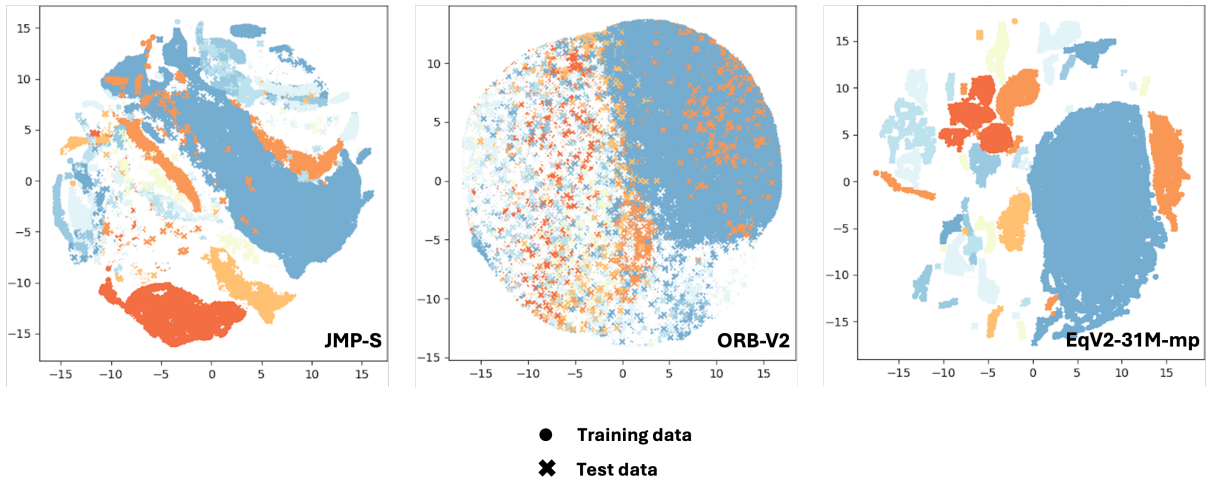

Figure 4: TSNE visualization of node representations on MP_E_form from four foundation models fine-tuned on the same dataset. The · symbols represent structures from the training set, while the × symbols indicate structures from the test set.

We further visualized the representation spaces of the fine-tuned models. We selected the MP_E_form dataset motivated by the fact that the fine-tuning results of the three models on this dataset showed notable differences (as detailed in Table 2). The visualization results reveal apparent similarities between the fine-tuned JMP-S and Equiformer models which cluster around element type, whereas ORB has no clear clustering similar to the non fine-tuned case. This pattern is consistent with JMP-S and Equiformer having similar MP_E_form accuracies, while ORB is significantly lower. However, the underlying link between the differences in the representation and the fine-tuning performance is still unclear and deserves further investigation, which can be easily facilitated with MatterTune.

## 4 Discussion and Conclusions

The MatterTune offers a flexible, generalizable framework that seamlessly integrated multiple atomistic FMs and supports tasks such as molecular dynamics simulations, materials property predictions, and materials discovery. MatterTune offers users a wide range of choices in data formats, model architectures, and training configurations. As a consequence of the modular design of MatterTune, users can freely mix and match these components according to their performance needs and specific requirements. Our experimental results to replicate the models' original reported performance on ambient water systems and JMP's Matbench experiments demonstrate that this unified approach to integrating atomistic FMs is both feasible and does not compromise model performance. We note that MatterTune is in active development and additional features will be planned in the future. The integration of additional newly developed atomistic FMs, as well as interfaces to more materials simulation software such as LAMMPs into MatterTune is ongoing. Furthermore, advanced fine-tuning procedures commonly seen in other contexts such as large language models will also be explored and implemented within the MatterTune toolkit.

To summarize, MatterTune is an effort to standardize and unify atomistic FMs while providing user-friendly interfaces for fine-tuning and applications. MatterTune also serves as a playground for experimenting and applying advanced fine-tuning algorithms to materials foundation models. By lowering the barrier to the use of atomistic FMs, we aim to make them broadly applicable across a wide range of materials science challenges, especially in materials simulations and informatics. Furthermore, we hope that the MatterTune platform can provide a foundation for exploring how to fine-tune atomistic FMs more effectively to meet the increasingly demanding requirements of materials science research.

# 5 Acknowledgements

This research used resources of the National Energy Research Scientific Computing Center, a DOE Office of Science User Facility supported by the Office of Science of the U.S. Department of Energy under Contract No. DE-AC02-05CH11231 using NERSC award BES-ERCAP0032102.